\documentclass[aps,prl,reprint,superscriptaddress]{revtex4-2}

\usepackage{graphicx} 
\usepackage{xcolor}
\usepackage{color,soul}
\usepackage{float} 
\usepackage{subcaption}
\usepackage{siunitx}
\usepackage{amsmath}
\usepackage{breqn}

\begin{document}


\title{Photonic time stretch fieldoscopy: single-shot electric field detection at\\near-petahertz bandwidth}

\author{Steffen Gommel}
\affiliation{Max Planck Institute for the Science of Light$,$ Staudtstra$\ss$e~2$,$ Erlangen$,$ 91058$,$ Germany}%
\affiliation{Friedrich-Alexander-Universit{\"a}t Erlangen-N{\"u}rnberg (FAU)$,$ Staudtstra$\ss$e~7$,$ Erlangen$,$ 91058$,$ Germany}
\affiliation{Graduate School in Advanced Optical Technologies (SAOT)$,$ Konrad-Zuse-Stra$\ss$e~3$,$ Erlangen$,$ 91052$,$ Germany}

\author{Kilian Scheffter}
\affiliation{Max Planck Institute for the Science of Light$,$ Staudtstra$\ss$e~2$,$ Erlangen$,$ 91058$,$ Germany}%
\affiliation{Friedrich-Alexander-Universit{\"a}t Erlangen-N{\"u}rnberg (FAU)$,$ Staudtstra$\ss$e~7$,$ Erlangen$,$ 91058$,$ Germany}
\affiliation{Graduate School in Advanced Optical Technologies (SAOT)$,$ Konrad-Zuse-Stra$\ss$e~3$,$ Erlangen$,$ 91052$,$ Germany}

\author{Andreas Herbst}
\affiliation{Max Planck Institute for the Science of Light$,$ Staudtstra$\ss$e~2$,$ Erlangen$,$ 91058$,$ Germany}%
\affiliation{Friedrich-Alexander-Universit{\"a}t Erlangen-N{\"u}rnberg (FAU)$,$ Staudtstra$\ss$e~7$,$ Erlangen$,$ 91058$,$ Germany}

\author{Anchit Srivastava}
\affiliation{Max Planck Institute for the Science of Light$,$ Staudtstra$\ss$e~2$,$ Erlangen$,$ 91058$,$ Germany}%
\affiliation{Friedrich-Alexander-Universit{\"a}t Erlangen-N{\"u}rnberg (FAU)$,$ Staudtstra$\ss$e~7$,$ Erlangen$,$ 91058$,$ Germany}

\author{Hanieh Fattahi}%
\email{hanieh.fattahi@mpl.mpg.de}
\affiliation{Max Planck Institute for the Science of Light$,$ Staudtstra$\ss$e~2$,$ Erlangen$,$ 91058$,$ Germany}%
\affiliation{Friedrich-Alexander-Universit{\"a}t Erlangen-N{\"u}rnberg (FAU)$,$ Staudtstra$\ss$e~7$,$ Erlangen$,$ 91058$,$ Germany}

\date{\today}

\begin{abstract}
Accessing the electric field of light with petahertz bandwidths in ambient air is a rapidly advancing frontier, essential for probing ultrafast dynamics driven by classical or quantum ultrashort pulses. Near-petahertz fieldoscopy has recently demonstrated sub-cycle access to light–matter interactions, enabling label-free spectro-microscopy of liquids and solids with unprecedented spatiotemporal resolution, detection sensitivity, and dynamic range. However, current implementations still rely on temporal scanning and averaging over many laser pulses. Here, we introduce photonic time-stretch fieldoscopy, enabling single-shot electric-field detection at near-petahertz frequencies. Numerical results demonstrate that integrating fieldoscopy with a nonlinear time lens enables the real-time acquisition of ultrashort optical waveforms with a detection bandwidth approaching petahertz. The resulting large temporal aperture and attosecond resolution allow direct single-shot detection of transient electric fields generated in solid or liquid samples. This concept opens new avenues for petahertz electronics, ultrafast spectro-microscopy, and the study of dynamic, non-repetitive optical phenomena.
\end{abstract}

\maketitle
\section{introduction}
Accessing the electric field of light with near-petahertz bandwidths in ambient air has become a major research frontier over the past decade~\cite{srivastava_near-petahertz_2024, keiber2016electro, alismail2020multi, Herbst_2022,zimin2021petahertz,cho2019temporal,park2018direct,korobenko2020femtosecond,sederberg2020attosecond,liu2021all,hui2022attosecond,bionta2021chip}, overcoming limitations inherent to vacuum-based attosecond streaking~\cite{hentschel2001attosecond, sommer2016attosecond, itatani2002attosecond, fattahi2016sub}. Fieldoscopy has emerged as a powerful approach for resolving sub-cycle light–matter interactions at near-petahertz bandwidths, providing temporal precision at tens-of-attoseconds and enabling simultaneous measurement of the electric field amplitude and phase with exceptional sensitivity and dynamic range~\cite{srivastava_near-petahertz_2024, zimin2025fieldoscopy}. Direct access to the electric field has, in turn, enabled access to the temporally gated material response relative to the driving pulse, allowing label-free spectro-microscopy of liquids and solids with unprecedented spatio-temporal resolution~\cite{Herbst_25, Jun2025_FemtosecondFieldoscopy, Srivastava2024_FieldResolvedFarField, Herbst2024_LabelFreeHyperspectralMicroscopy, Scheffter2025_MulitmodalFieldResolvedSpectroscopy, Scheffter_25_KerrAmplificationLiquidThinFilm}. 

In Fieldoscopy, the electric field of a signal pulse is resolved by scanning a temporally shorter gate pulse across the signal waveform. At each delay step, the nonlinear interaction between the gate and signal pulses encodes the instantaneous electric field. Despite its outstanding temporal resolution, the technique relies fundamentally on sequential time-domain scanning and signal averaging over multiple laser pulses. This dependence imposes significant limitations on acquisition speed in both imaging and spectroscopic measurements, leading to increased dwell times and hindering investigations of fragile samples or rare phenomena~\cite{jun2024nonlinear, rasputnyi2024high, sennary2025attosecond}. 

The dispersive Fourier transform (DFT) has proven highly effective for single-shot detection of optical intensity waveforms by mapping spectral components into the time domain~\cite{kawai2020time, goda2013dispersive, mahjoubfar2017time, nakamura2024broadband, Jannson:83, goda:09, Runge:15, shoshin2025mid, Krupa:17, Runge:14, Herink:17, zhou:2022, Salem:13, goda:13}. Integrating electro-optic sampling with DFT or compressed-sensing strategies has further enabled single-shot or real-time measurements of the electric field at terahertz frequencies~\cite{scheffter2024compressed,Szwaj:16, Couture:23,  couture2023single}. Nonetheless, achieving single-shot detection of the electric field of light at near-petahertz frequencies remains challenging, as current approaches rely on reconstruction algorithms, multi-shot operation, or high-energy laser pulses, and lack the detection sensitivity and dynamic range required to resolve faint samples' response ~\cite{ryczkowski:18, Tikan:18, truong2025scanless, yeom2024single}. 

In this work, we present photonic time-stretch fieldoscopy (PTF) for real-time access to the electric field of optical ultrashort pulses. By combining fieldoscopy with a nonlinear time lens, PTF enables single-shot electric-field detection with bandwidths extending into the petahertz range. The nonlinear time lens is configured analogously to a spatial 2-f imaging system, where the time lens acts as a temporal Fourier transformer, mapping the signal pulse onto the sum frequency spectrum of the signal and focal pulses. The sum frequency pulse is then combined with the focal pulse in a field detection scheme. Eventually, by performing balanced detection using either dispersive Fourier transform or angularly dispersed spectral balancing, PTF enables direct, single-shot detection of the signal electric field. 

By converting ultrafast optical waveforms into a time-stretched domain, PTF eliminates the need for temporal scanning, enabling each laser pulse to be recorded in real-time. Through numerical simulations, we establish the fundamental requirements for implementing PTF and analyze the intrinsic limits that determine its temporal resolution, detection bandwidth, and detection dynamic range. This approach lays the foundation for extending attosecond-scale field measurements into the single-shot regime, opening new opportunities for observing sub-cycle, non-repetitive light–matter interactions.

    \begin{figure*}
    \centering
    \includegraphics{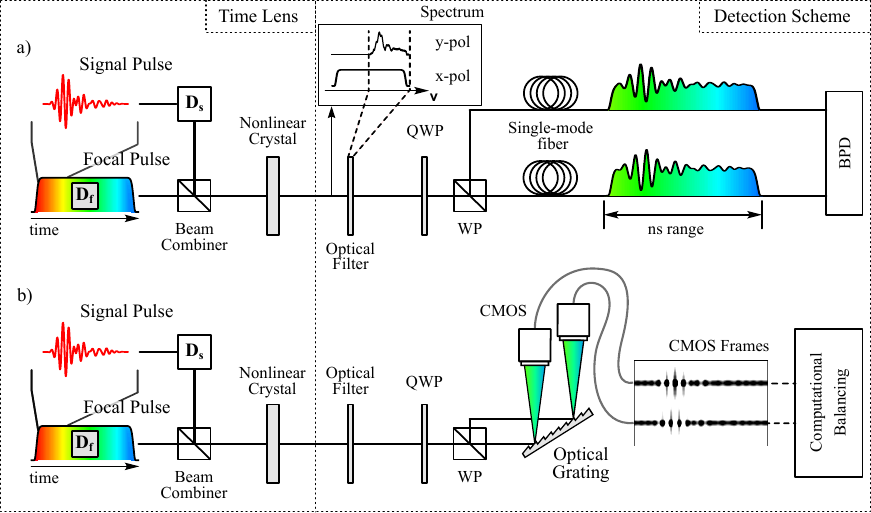}
    \caption{Photonic time-stretch Fieldoscopy (PTF) for single-shot detection of the electric field of light at near-petahertz bandwidth. The field detection \cite{srivastava_near-petahertz_2024} is performed under the time-lens condition. Two approaches can be employed to detect the signal carrying the field information: a) utilizing photonic time stretch and balanced photodetection at GHz sampling rates, b) utilizing polarization-resolved spectrometry, CMOS sensor, and computational balancing. $\text{D}_{\text{s}}$: signal dispersion; $\text{D}_{\text{f}}$: focal dispersion; QWP: quarter waveplate; WP: Wollaston prism; BPD: balanced photodiode.}
    \label{fig:PTF_fieldoscopy}
    \end{figure*}

\section{Photonic time-stretch Fieldoscopy: the concept}

The PTF concept is illustrated in Fig.~\ref{fig:PTF_fieldoscopy}. The objective is to detect the electric field of a signal pulse that may carry a sample response, such as a molecular vibrational signature. In liquids or solids, such responses can persist from hundreds of femtoseconds to several picoseconds~\cite{srivastava_near-petahertz_2024, Herbst_25}. For accurate mapping, the focal and signal pulses must be temporally synchronized and are linearly chirped to satisfy the time-lens condition. Analogous to placing an object at the focal plane of a spatial $2f$ imaging system, setting $D_s = -D_f$ positions the signal field at the focus of the time lens. To resolve the electric field, the stretched signal and focal pulses are spatio-temporally overlapped in a nonlinear crystal for sum-frequency generation (SFG). Here, the time lens maps the temporal intensity profile of the signal pulse onto the SFG spectrum.

To resolve the electric field, the SFG signal must spectrally overlap with the focal pulse, as illustrated in the inset of Fig.\ref{fig:PTF_fieldoscopy}~a) and has been shown in \cite{srivastava_near-petahertz_2024, keiber2016electro}. By adjusting the temporal delay between the signal and focal pulses, the central frequency of the SFG spectrum can be tuned. The portion of the SFG spectrum that overlaps with the residual focal pulse is then spectrally filtered. We refer to this spectral region as the spectral overlap bandwidth (SOB). The field information encoded in the SOB is carried in two orthogonal polarizations. These polarizations are spatially separated using a quarter-wave plate and a Wollaston prism. PTF is then achieved through single-shot balanced detection of the two orthogonal polarizations. Two approaches can be employed:

i) Time-domain (dispersive) detection: Both beams are stretched to perform frequency-to-time mapping of the SOB, as shown in Fig.~\ref{fig:PTF_fieldoscopy}~a). Balanced detection of the stretched outputs using GHz-bandwidth photodiodes enables direct single-shot measurement of the signal pulse’s electric field.

ii) Spectral-domain detection: Alternatively, the spectra of the two orthogonally polarized beams can be imaged onto a CMOS sensor using a grating, as shown in Fig.~\ref{fig:PTF_fieldoscopy}~b). The CMOS records the two spectral traces, after which computational balancing of corresponding pixel pairs is performed to access the field information~\cite{Black:15, Mamaikin_22}.

\section{Numerical study}

\begin{figure}[t]
    \centering
    \includegraphics[width=1\columnwidth]{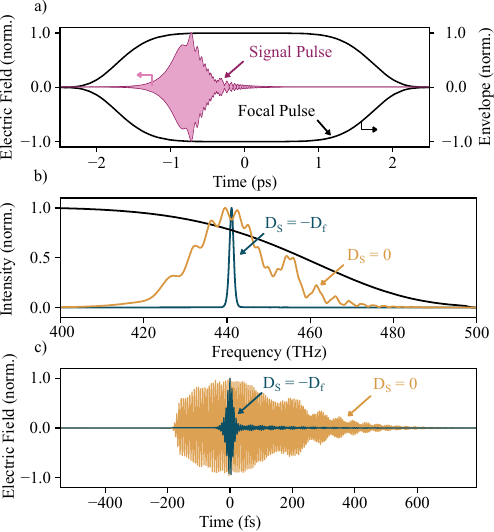}
    \caption{a) The temporal overlap between the signal pulse (purple electric field) stretched by $D_s$=\SI{-2110}{\femto\second\squared} and the focal pulse (black envelope) stretched by $D_f$=\SI{2110}{\femto\second\squared}. b) Sum frequency spectrum resulting from the interaction of the stretched focal pulse ($D_f$=\SI{2110}{\femto\second\squared}) and the FTL signal pulse ($D_s = 0$) is shown in the orange curve, while the teal curve corresponds to the sum frequency spectrum in which the signal pulse is stretched by $D_s = -D_f$. The black curve indicates the portion of the focal-pulse spectrum that overlaps with the sum-frequency spectrum. c) The corresponding resolved electric fields of the signal pulse from the generated sum frequency spectra in panel (a). The resolved electric field under the time-lens condition $D_s = -D_f$ (teal) matches the original signal pulse. In contrast, the electric field retrieved for $D_s = 0$ (orange) exhibits an additional temporal chirp that is absent in the original signal pulse. The SOB spans from 425\,THz to 500\,THz.} 
    \label{fig2}
    \end{figure}

Numerical simulations were performed using an open-source nonlinear wave equation solver~\cite{Karpowicz:github:25, Karpowicz:23}. The simulation parameters were chosen to match closely the values reported in \cite{srivastava_near-petahertz_2024}. Further details of the simulation are provided in the endmatter. Fig.~\ref{fig2}~a) shows the electric field of the signal pulse and the envelope of the focal pulse, both stretched to satisfy the time-lens condition. Fig.~\ref{fig2}~b) shows the generated SFG spectrum by the interaction of the signal and focal pulses. The central frequency of the sum frequency spectrum can be adjusted by varying the temporal overlap between the signal and focal pulses. The temporal overlap was adjusted to achieve a SOB spanning from 425\,THz to 500\,THz. The orange curve corresponds to the sum frequency spectrum where the focal pulse is stretched ($D_f = \SI{2110}{\femto\second\squared}$), and the signal pulse remains Fourier-transform-limited ($D_s = 0$). The teal curve represents the sum frequency spectrum in which the signal pulse is stretched to satisfy the time-lens condition 
($D_s = -D_f$).

As illustrated in the concept in Fig.~\ref{fig:PTF_fieldoscopy}, the SFG spectrum is then numerically propagated through a bandpass filter (SOB range of \SI{425}{\tera\hertz} to \SI{500}{\tera\hertz}), a quarter-wave plate, and a Wollaston prism. Eventually, the intensity spectra of orthogonal polarization axes are subtracted to simulate the balanced detection. Fig.~\ref{fig2}~c) shows the retrieved electric field of the signal pulse for the two dispersion cases. Under the condition of $D_s = -D_f$, the resolved electric field (teal curve) is identical to the original signal pulse we aim to measure (see Fig.~\ref{SI_fig1} in endmatter). In contrast, for $D_s = 0$ the resolved electric field exhibits additional temporal chirp, arising from the fact that the time-lens condition is not satisfied. 
 
    \begin{figure}[t]
    \centering
    \includegraphics[width=1\columnwidth]{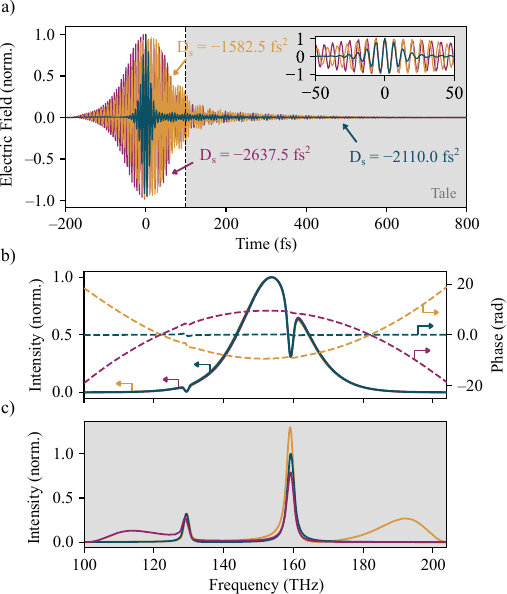}
    \caption{a) Overcompensating or undercompensating the focal dispersion by applying $\text{D}_s = \SI{-2637.5}{\femto\second\squared}$ (purple curve) or $\text{D}_s = \SI{-1582.5}{\femto\second\squared}$ (orange curve) results in a down-chirped or up-chirped retrieved electric field of the signal. The teal curve shows the time-lens condition for $\text{D}_s = \SI{-2110}{\femto\second\squared}$. The inset shows a magnified view of all three resolved electric fields at time zero. b) The corresponding spectra and spectral phases. c) The spectra of the temporally gated resolved signal pulse. The gated region is shown in gray in panel a).} 
    \label{fig:3}
    \end{figure} 
    
Fig.~\ref{fig:3}~a) illustrates how overcompensating or undercompensating the focal dispersion results in a down-chirped or up-chirped retrieved electric field of the signal. The corresponding retrieved spectra are displayed in Fig.~\ref{fig:3}~b), along with their spectral phase. The group delay dispersion extracted from the retrieved spectral phase for both cases agrees with the sum of the applied focal and signal dispersions to within an error of $\text{D}_{error} = \SI{3.5}{\femto\second\squared}$, which corresponds to the numerical accuracy of the simulation. This indicates that, when the input dispersion values of the PTF system are known, the phase of the retrieved signal field can be corrected computationally without spectral distortion, as long as the retrieved spectrum lies within the SOB.

To assess the accuracy of the PTF in resolving absorption features located in the trailing edge of the signal pulse, the resolved electric field is temporally gated from \SI{100}{\femto\second} to \SI{800}{\femto\second} and Fourier transformed (see Fig.~\ref{fig:3}~c)). Under the time-lens condition, the Lorentzian-shaped absorptions are accurately reconstructed. In contrast, the up- and down-chirped cases show additional spurious frequency components, originating from dispersive leakage of the main pulse into the gated temporal region. Nevertheless, the true spectral features remain clearly distinguishable.

\begin{figure}[t]
    \centering
    \includegraphics[width=1\columnwidth]{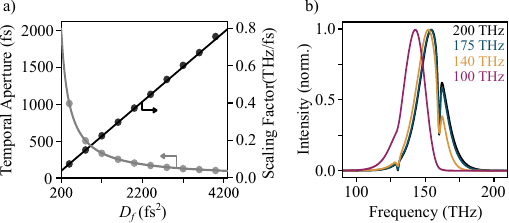}
    \caption{a) The PTF temporal aperture and scaling factor as a function of $D_f$, under the time lens condition $D_f = -D_s$. b) Signal spectra retrieved for varying focal-pulse bandwidths centered at \SI{345}{\tera\hertz} under the time-lens condition. The SOB spans from 425\,THz to 500\,THz.} 
    \label{fig4}
    \end{figure} 
    
To accurately resolve the electric field of the signal pulse after resonant interaction with the medium, the observable time window must span several hundred femtoseconds. This window, referred to as the temporal aperture of the system, increases with the dispersion of the focal pulse $D_f$ in PTF \cite{Kolner:00}. However, increasing the temporal aperture inevitably reduces the sampling bandwidth per unit time. To quantify this trade-off, we define a figure of merit, the \textit{scaling factor}, as the ratio of the SOB to the temporal aperture. A larger scaling factor corresponds to a narrower effective bandwidth of the filtered SFG signal. This can either demand very high group-delay dispersion, as in the design of Fig.\ref{fig:PTF_fieldoscopy}~a), or reach the spectrometer’s resolution limit for the filtered SFG bandwidth, as in the design of Fig.\ref{fig:PTF_fieldoscopy}~b).
 
Fig.~\ref{fig4}~a) shows the dependence of the temporal aperture and scaling factor on the focal dispersion $D_f$, for the case in which the time-lens condition $D_s = -D_f$ is satisfied, assuming an SOB spanning \SI{425}{\tera\hertz} to \SI{500}{\tera\hertz}. As the focal dispersion increases, the temporal aperture of the PTF expands, and the scaling factor correspondingly decreases. The maximum resolvable temporal aperture is strongly constrained by the temporal or spectral resolution of the spectrometer or the DFT-based balanced detection system.

Fig.~\ref{fig4}~b) shows the quality of the resolved electric field for various bandwidths of the focal pulse centered at \SI{345}{\tera\hertz}, constrained by the requirement to maintain spectral overlap between the SFG and the focal pulse. Using a second-order nonlinear process imposes two principal conditions on the central frequency and bandwidth of the focal and signal pulses. First, the focal pulse must be centered at a higher frequency than the signal pulse, yielding the condition $\omega_0 > \Omega_0$. Second, to ensure temporal overlap between the two homogeneously stretched pulses, and to maintain sufficient spectral overlap between the nonlinearly converted pulse and the focal pulse, the following condition must be satisfied: $\Omega_0 + \frac{\Delta\Omega}{2} < \Delta\omega,$ where $\Delta\Omega$ and $\Delta\omega$ are the bandwidths of the signal and focal pulses, respectively.
    
\section{Discussion}

We have introduced the concept of photonic time-stretch fieldoscopy (PTF), establishing a framework for extending electric field measurements with attosecond temporal resolution into the single-shot regime. PTF enables scan-free detection of the electric field of ultrashort pulses with bandwidths approaching the petahertz range by integrating fieldoscopy with a nonlinear time lens. The numerical study showed that the system’s focal dispersion governs both the temporal aperture and the resolution requirements of PTF. In contrast to conventional field detection techniques, PTF doesn't require carrier-envelope-phase (CEP) stability of the signal pulse, or temporal compression of the ultrashort gate pulses to access the sub-cycle electric field information. The maximum acquisition rate of PTF is limited by the repetition rate of the driving laser and the acquisition time of the detectors. 

The nonlinear time lens can also be implemented using difference-frequency generation (DFG) or four-wave mixing (FWM). Because these processes involve conjugation of one of the interacting fields, they require the signal and focal dispersions to share the same sign~\cite{Kolner:00, Salem:08, schroeder:10, Pasquazi:12}. A DFG-based time lens can be particularly advantageous experimentally, as the generated signal lies in the near-infrared, where optical fibers offer a more favorable stretching-to-attenuation ratio~\cite{Sulzer:20, Kolner:00, Hanoun:24}. Alternatively, as proposed here, angularly dispersive or diffractive optics may be used to extract the spectral information, eliminating the need for extremely long fibers and avoiding the practical challenges of precisely matching dispersion across multiple fiber paths. Polarization-resolved spectra can be obtained by detecting both polarization components, either using a CMOS sensor or two synchronized spectrometers, followed by electronic or digital balancing.

A further consideration is the impact of higher-order dispersion when stretching broadband pulses. As discussed in the end matter, incorporating third-order dispersion (TOD) results in the nonlinear time to frequency mapping, an effect that must be accounted for and that is expected to persist for even higher dispersion orders. To prevent third-order distortions in a SFG time lens, the TOD of both the signal and focal pulses must match in both sign and magnitude, a condition that is generally favorable because dispersive media in the relevant spectral range typically exhibit positive TOD.

PTF offers a promising route for real-time ultrafast spectroscopy with high detection sensitivity and dynamic range. Its large temporal aperture enables direct access to the material response, allowing real-time probing of molecular dynamics that unfold outside the main excitation pulse. This capability opens new opportunities for studying ultrafast processes under rapidly changing conditions, investigating fragile samples with reduced dwell times~\cite{jun2024nonlinear}, advancing petahertz electronics \cite{heide2024petahertz}, and studying rare or non-repetitive classical or quantum events \cite{rasputnyi2024high, sennary2025attosecond} in the near-infrared and visible spectral ranges.

\section*{DECLARATIONS}
This work was supported by research funding from the Max Planck Society and funding from the ERC Consolidator Grant ID: 101125670. The authors declare no competing interests. The data supporting this study's findings are available from the corresponding author upon reasonable request.

\bibliography{main}

\renewcommand{\thefigure}{E\arabic{figure}}  
\setcounter{figure}{0}                       
\section{end matter}

    \begin{figure}[H]
    \centering
    \includegraphics[width=1\columnwidth]{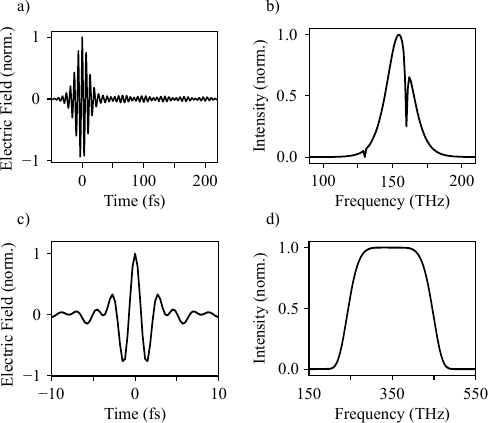}
    \caption{a) The electric field of the signal pulse. b) The signal spectrum. c) The electric field of the focal pulse. d) The focal spectrum.} 
    \label{SI_fig1}
    \end{figure} 

\subsection{Simulation parameters}
Numerical simulations were performed using the UPPE model of an open-source nonlinear wave equation solver, configured to operate under the slowly varying wave approximation~\cite{Karpowicz:github:25, Karpowicz:23}. The initial focal and signal fields (see Fig.\ref{SI_fig1}) are defined to be linearly polarized along the x- and y-axes, respectively. The focal pulse is modeled using a sixth-order super-Gaussian spectrum centered at \SI{345}{\tera\hertz} with a \SI{200}{\tera\hertz} full-width-at-half-maximum (FWHM) bandwidth, corresponding to a Fourier-transform-limited pulse duration of \SI{3}{\femto\second} (FWHM). The signal pulse is modeled with a sech-shaped envelope centered at \SI{155}{\tera\hertz}, yielding a Fourier-transform-limited pulse duration of \SI{14}{\femto\second} (FWHM). We assume that the signal pulse has previously undergone resonant absorption. The system is modeled as two isolated Lorentzian oscillators at \SI{130}{\tera\hertz} and \SI{160}{\tera\hertz}, with dephasing times of \SI{0.5}{\pico\second} and \SI{0.25}{\pico\second}, respectively. 

To model the SFG between the signal and focal pulses in a $\chi^{(2)}$ nonlinear crystal, we use a \SI{20}{\micro\meter}-thick BBO crystal \cite{Shoji:99, Bache:13,zhanng:00} with a phase-matching angle of $\theta = \SI{25}{\degree}$. The crystal is configured for type-II phase matching, generating an SFG pulse polarized along the y-axis. The energies of both the focal and signal pulses are set to \SI{160}{\pico\joule}. The focal pulse and the signal pulse are focused to \SI{20}{\micro\meter}, and \SI{30}{\micro\meter} (at $\frac{1}{e^2}$) at the nonlinear crystal. The quarter-wave plate and a Wollaston prism are modeled as a single rotatable Wollaston prism.
For positive $D_f$, the frequency axis is related to the time axis by
    \begin{equation}
    \tau(\omega) = -D_f \left( \omega - \omega_0 - \Omega_0 \right) \,,
    \label{eq:mapped_time}
    \end{equation}
where $\tau(\omega)$ is defined by the focal dispersion, $D_f = -D_s$, $\omega_0$ and $\Omega_0$ are the central frequency of the focal pulse and the signal pulse, respectively.

\subsection{Homogeneous Stretching}
Considering homogeneous stretching described by the Taylor expansion at the central frequency up to the second order while neglecting dispersive propagation after the detection crystal, $\Phi(\omega, \Omega)$ simplifies to
\begin{eqnarray}
\Phi(\omega, \Omega) =& \, \phi_{s}(\Omega) + 
\phi_{s,1}(\Omega) + \phi_{s,2}(\Omega) + \phi_{f,1}(\omega - \Omega) \nonumber \\ 
& + \phi_{f,2}(\omega - \Omega)
- \phi_{f,1}(\omega) - \phi_{f,2}(\omega) \,.
\label{eq:phase}
\end{eqnarray}
Here $\phi_s(\Omega)$ denotes the unknown phase of the signal field prior to stretching, whereas in the remaining known Taylor expansion phase terms, the first index assigns the field (either focal or signal field) and thereby the central frequency, while the second index assigns the expansion order of the phase term. To achieve temporal imaging, before the nonlinear interaction, the group delay dispersion (GDD) of both beams must match, but with opposite signs. Adding the condition that both fields are temporally overlapped in the nonlinear crystal and considering positive chirping of the focal pulse, eq.~\ref{eq:phase} simplifies to
\begin{subequations}
\label{eq:phase_group}
\begin{equation}
\Phi(\omega, \Omega) = \phi_{s}(\Omega) + \Omega \, \tau(\omega) - \Omega_0 \left( \phi_{1} + \frac{1}{2} \phi_2 \Omega_0 \right) \,,
\label{subeq:phase2}
\end{equation}
\begin{equation}
\tau(\omega) = -\phi_2 (\omega - \omega_0 - \Omega_0) \,.
\label{subeq:mapped_time}
\end{equation}
\end{subequations}
Here, $\phi_1 = \frac{L_f}{v_{f,g}} = \frac{L_s}{v_{s,g}}$, with dispersive lengths $L_{s,f}$ and propagation in material or air chosen to achieve temporal overlap at the nonlinear crystal.  
The conversion of the frequency axis of the balanced intensity spectrum into a time vector, as indicated by eq.~\ref{subeq:mapped_time}, depends on the focal dispersion with $D_f=-D_s=\phi_2$, as well as on $\Omega_0$ and $\omega_0$, the central frequencies of the signal and focal fields, respectively. The required focal dispersion with a given temporal aperture $\Delta \tau$ and SOB is calculated as
\begin{equation}
D_f = \left| \frac{\Delta \tau}{SOB} \right| \,.
\label{eq:req_D_f}
\end{equation}

The third term on the right-hand side of eq.~\ref{subeq:phase2} represents a constant phase offset, independent of $\omega$. 
This offset corresponds to a shift of the carrier-envelope phase (CEP) of the detected signal field.  



\subsection{Influence of TOD}
Considering the TOD introduced by the stretcher before the time lens, the sign and magnitude of the signal and focal field TOD must satisfy the condition $\text{TOD}_f = \text{TOD}_s = \phi_3 \, [fs^3]$ to prevent third-order distortions in the retrieved field.  

Using the same formalism as presented in eq.~\ref{eq:phase}, but now expanding up to third order, the phase becomes  
\begin{subequations}
\label{eq:phase_group_TOD}
\begin{equation}
\begin{split}
\Phi(\omega, \Omega) &= \phi_{s}(\Omega) + \Omega \tau(\omega)  
+ \Omega^2 \frac{1}{3}\phi_3(\omega-\omega_0-\Omega_0) \\
&\quad - \Omega_0 \left( \phi_{1} + \frac{1}{2} \phi_2 \Omega_0 \right)
- \frac{1}{6}\phi_3\Omega^3_0
\end{split}
\label{subeq:phase_TOD}
\end{equation}
\begin{equation}
\begin{split}
\tau(\omega) =& -\phi_2 (\omega-\omega_o-\Omega_0) \\
&\quad - \frac{1}{3}\phi_3\left[(\omega-\omega_0)^2 - \Omega_0^2\right]
\end{split}
\label{subeq:mapped_time_TOD}
\end{equation}
\end{subequations}

The third term on the right-hand side of eq.~\ref{subeq:phase_TOD} introduces a chirp to the retrieved waveform. As before, the following terms correspond to a constant phase offset and are therefore less relevant. With TOD present, the newly defined time vector in eq.~\ref{subeq:mapped_time_TOD} now has a quadratic dependence on $\omega$, introducing a distortion to the time axis that must be accounted for.


\end{document}